\documentclass[authoryear,3p,11pt]{elsarticle}
\usepackage{amsmath,amssymb,mathtools,mathrsfs}

\usepackage{ulem}
\usepackage{cancel}
\usepackage{float}
\usepackage{bm}

\usepackage[pdftex, pdftitle={Article}, pdfauthor={Author}]{hyperref} 
\usepackage[dvipsnames]{xcolor}
\usepackage{booktabs}
\usepackage{multirow}

\usepackage[symbol]{footmisc}

\journal{Elsevier}
\usepackage[nodayofweek,level]{datetime}
\newdate{date}{today}
\date{\displaydate{date}}

\begin{document}

\begin{frontmatter}

\title{Inverse design of a magneto-elastica for shape-morphing}

\author{JiaHao Li$^{1,\dagger}$, Yingchao Zhang$^{2,\dagger}$, Weicheng Huang$^{3,\dagger}$, Shenghao Ye$^{4}$, HengAn Wu$^{1,}$\footnote[1]{\textit{Corresponding Author: wuha@ustc.edu.cn (H.W.)}}, \\ Dominic Vella$^{4,}$\footnote[1]{\textit{Corresponding Author: dominic.vella@maths.ox.ac.uk (D.V.)}}, and Mingchao Liu$^{{5,}}$\footnote[1]{\textit{Corresponding Author: m.liu.2@bham.ac.uk (M.L.)}} }

\address{$^{1}$\:State Key Laboratory of Nonlinear Mechanics, Department of Modern Mechanics, University of Science and Technology of China, Hefei 230027, People’s Republic of China\\
$^{2}$\:AML, Department of Engineering Mechanics, Tsinghua University, Beijing, China, 100086\\
$^{3}$\:School of Engineering, Newcastle University, Newcastle upon Tyne NE1 7RU, UK \\
$^{4}$\:Mathematical Institute, University of Oxford, Woodstock Rd, Oxford, OX2 6GG, UK\\
$^{5}$ Department of Mechanical Engineering, University of Birmingham, Birmingham, B15 2TT, UK \\
$^{\dagger}$These authors contributed equally to this work.
}

\begin{abstract}
Slender magnetic elements provide a versatile platform for programmable shape-morphing under remote magnetic actuation. However, a general and physically interpretable framework for the inverse design of a `magneto-elastica' under prescribed boundary conditions remains lacking.
In this work, we develop an explicit analytical formulation for the inverse design of a magneto-elastica based on the integral form of the moment equilibrium equations. Compared with purely local differential formulations, this integral approach makes the global constraints imposed by boundary conditions and moment balance explicit, and enables a systematic characterization of the feasible design space.
We identify the key dimensionless parameters that govern the competition between magnetic torques and elastic restoring moments and show that the applied boundary conditions are an essential ingredient. We obtain closed-form solutions for the beam tapering profiles required to generate desired actuated shapes in the cases of clamped–free and clamped–clamped configurations; in the latter case, this includes analytical expressions for the boundary reactions.
The formulation recovers the classical inverse elastica in the absence of magnetic fields and reveals a linear scaling between curvature deviation and magnetic mismatch. A tessellation strategy based on stiffness tailoring is further proposed for the design of discretized morphing surfaces. The theoretical predictions are validated against discrete elastic rod simulations and experiments across representative geometries.
This work establishes a consistent analytical framework for the inverse design of a magneto-elastica and provides new insight into magnetically-induced shape programming in slender structures.
\end{abstract}

\begin{keyword}
Magneto-elastica \sep Shape-morphing \sep Inverse design \sep Analytical framework
\end{keyword}

\end{frontmatter}

\section{Introduction}
\label{sec:Introduction}

Shape-morphing structures that reconfigure under external stimuli provide a route to adaptivity and multifunctionality in mechanical systems, enabling a single device to switch between configurations, mechanical responses, and modes of interaction with its environment \citep{sharonMechanics2010,oliver2016morphing,sydney2016biomimetic,van2017growth,dudek2025shape}. Such capabilities are central to applications including compact-to-deployed transformations \citep{gattas2015geometric}, reconfigurable locomotion and manipulation \citep{sun2023embedded}, and conformal interfacing with complex and time-varying surfaces \citep{qamar2018hci}. Common strategies to achieve programmable morphing include geometric programming through architected layouts, material programming through built-in eigenstrain or intrinsic curvature, and field-driven actuation using remotely applied stimuli \citep{yang2023morphing}. In many cases, complex three-dimensional (3D) shapes are generated by designing low-dimensional building blocks, such as beams, ribbons, and plates, whose kinematics allow large deformations at relatively low energetic cost \citep{liu2020tapered,bo2023mechanically,li2026inverseElastica}.

Among field-driven actuation mechanisms, magnetic actuation has emerged rapidly in recent years due to its remote and tunable nature \citep{bastola2021shape}. By embedding magnetic particles within elastic matrices, structures can undergo large and reversible deformations under externally applied magnetic fields \citep{kim2022magnetic}. Their mechanical response is governed by the competition between elastic bending and magnetically induced body torques, leading to rich nonlinear behaviors and a broad range of accessible configurations \citep{zhao2019mechanics}. Magnetically actuated slender structures have therefore attracted significant attention as a platform for programmable shape-morphing \citep{lum2016shape,hu2018multimode,wang2021evolutionary}. In particular, one-dimensional (1D) magnetic structures, such as beams and rods, are especially attractive thanks to their ability to undergo large, controllable deformations \citep{yanComprehensive2022,sanoKirchhofflike2022,zhangNoncontact2023}. Such characteristics make them fundamental building blocks for morphing systems and have enabled applications in areas such as continuum robotics and minimally invasive devices, where flexible magnetic elements are used for navigation and manipulation \citep{kim2019ferromagnetic,kim2022magnetic,dreyfusDexterous2024,tong2025real,wang2026magnetically}.

Inverse design plays a central role in exploiting these capabilities: the goal is to determine the structural parameters or actuation that will produce prescribed shapes. For slender structures, inverse design has been studied extensively in the context of the (non-magnetic) elastica, including strain-programmed bilayer ribbons \citep{chen2011tunable,armon2011geometry,li2025biomimetic,levinHierarchy2021}, buckling-driven rods with tailored intrinsic curvature and boundary conditions \citep{derouet2010stable,derouet2013inverse,bertails2018inverse,qin2022bottom,li2026inverseElastica}, and stiffness-programmed structures \citep{liu2020tapered,fan2020inverse,hafner2021design,zhang2022shape,hafner2023design}. For the stiffness-programmed structures, a particularly convenient way of varying the stiffness spatially is to taper the strip width. However, many existing stiffness-programming approaches are formulated through local differential elastica equations, which relate curvature, loading, and stiffness pointwise. Although powerful, such formulations do not by themselves expose the global compatibility and moment-balance constraints imposed by prescribed boundary conditions. This limitation becomes particularly important when moving from the classical elastica to the magneto-elastica, where the magnetically induced moments must be compatible with the entire target shape \citep{wang2020hard}.

The inverse design of the magneto-elastica introduces additional challenges beyond those encountered in classical elastica \citep{lum2016shape,wang2021evolutionary,wang2023analytical,li2025programmable}. Magnetic actuation enters the equilibrium equations as a distributed torque density, giving rise to intrinsically nonlocal contributions to the internal moment \citep{wang2020hard}. At the same time, the governing equations remain strongly nonlinear because of the geometric coupling between curvature and rotation \citep{sanoKirchhofflike2022}. As a result, existing inverse design approaches based on the differential form of the elastica equations \citep{liu2020tapered,fan2020inverse} do not directly reveal global constraints on admissible curvature and rotation fields, making it difficult to characterize the feasible design space under prescribed boundary conditions.

In this work, we develop an explicit analytical framework for the inverse design of the shape of the magneto-elastica. Our approach is based on the integral form of the moment equilibrium equations and focuses on the strip tapering required to generate prescribed actuated shapes under magnetic loading; this tapering is more amenable to manufacturing than changing the spatial distribution of magnetic moments. This integral formulation directly incorporates the global constraints imposed by boundary conditions and magnetic actuation, yielding explicit conditions on admissible curvature and rotation fields and enabling a systematic characterization of the feasible design space. 

By introducing appropriate dimensionless groups, we identify the key parameters governing the competition between magnetic torques and elastic restoring moments. Closed-form solutions are derived for both clamped--free and clamped--clamped configurations, including analytical expressions for the boundary reactions in the latter case. Our framework naturally recovers the classical inverse elastica in the absence of magnetic actuation under clamped--clamped boundary conditions~\citep{liu2020tapered}. Motivated by the ability of the classical elastica to fabricate 3D surfaces with rotational symmetry by varying strip width and stiffness~\citep{liu2020tapered, zhang2022shape}, we show how this can be achieved in the magneto-elastic case. In each case, our theoretical predictions are validated through comparisons with discrete elastic rod simulations and experiments across representative examples.

The remainder of the paper is organized as follows. 
Section \ref{sec:analy_inver_design} formulates the integral equilibrium equations and develops the inverse-design framework. The general theoretical framework is presented in Sections \ref{sec:Integral_equations} and \ref{sec:Dimensionless_formulation}. Analytical solutions under different boundary conditions are given in Sections \ref{sec:Clamped_free_BC} and \ref{sec:Clamped_clamped_BC}, followed by a discussion of the connection to the classical elastica in Section \ref{sec:Reduction_to_classical}, the tessellation strategy in Section \ref{sec:Tessellation}. Experimental and numerical validations are presented in Section \ref{sec:res_dis}. Section~\ref{sec:extension} discusses the extension of the present theory to elastica with spatially non-uniform magnetization directions, and conclusions are drawn in Section \ref{conclusion}.

\section{Analytical inverse design framework}
\label{sec:analy_inver_design}

In this section, we develop an analytical framework for the inverse design of a magneto-elastica subject to a uniform magnetic field. The formulation is based on the integral form of the moment equilibrium equations, which provides a global description of the internal force–moment balance. The governing relations are then cast in dimensionless form to identify the key parameters controlling the competition between magnetic and elastic effects.

The framework is subsequently applied to inverse design (i.e.~determining the tapering that yields a desired 3D shape) under two canonical boundary conditions (BCs): clamped--free and clamped--clamped configurations. Finally, an alternative design strategy based on tessellation is introduced, enabling the extension of the framework to discretized morphing structures.

\subsection{Integral equations of moment equilibrium}
\label{sec:Integral_equations}

\begin{figure}[h]
    \centering
    \includegraphics[width=\textwidth]{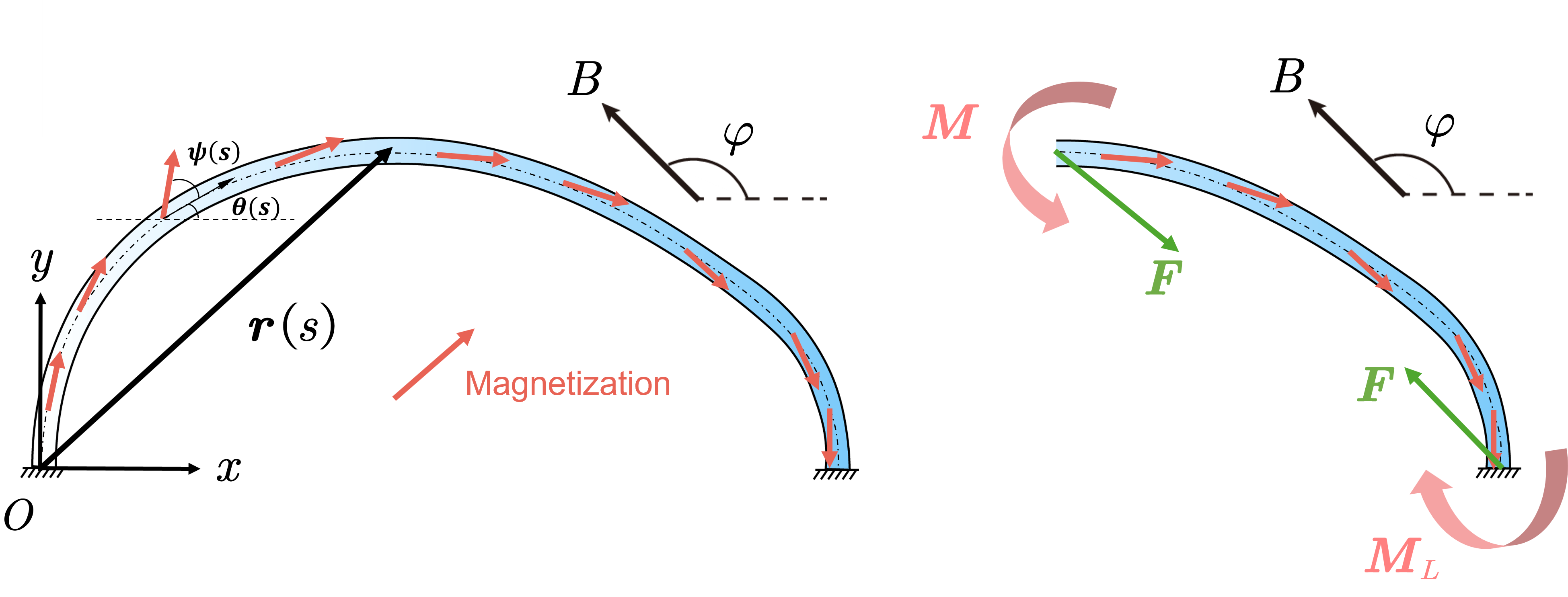}
    \caption{\textbf{Schematic of a magneto-elastica under a uniform external magnetic field.}
    (a) Geometry and kinematics of the beam. The beam centerline is parameterized by the position vector $\mathbf{r}(s)$, where $s$ represents the arc length measured from the origin $O$. The angle $\theta$ denotes the inclination of the beam tangent with respect to the horizontal direction and $\psi$ denotes the angle between the magnetization direction and the tangential direction of the elastica. The beam is subjected to a uniform magnetic field $\mathbf{B}$ inclined at an angle $\varphi$. Red arrows indicate the distribution of magnetization along the beam. 
    (b) Free-body diagram of a beam segment. $\mathbf{F}$ and $\mathbf{M}$ denote the internal force and moment, respectively, and $\mathbf{M}_L$ is the reaction moment at the clamped boundary.}
    \label{fig:1}
\end{figure}

We consider a slender elastic beam of length $L$ with a non-uniform cross-section (Fig.~\ref{fig:1}(a)), clamped at one or both ends and subjected to a uniform magnetic field of magnitude $B$ and orientation $\varphi$.
The beam is assumed to have a constant magnetization magnitude $M$, while the magnetization direction is prescribed relative to the local tangent. Specifically, the magnetization vector forms an angle $\psi(s)$ with the local tangent, where $s \in [0, L]$ is the arc-length coordinate measured from the clamped end ($s=0$).
The mechanical response is governed by the balance between elastic bending moments and magnetically induced torques. Owing to its slenderness, the beam's deformation is dominated by bending, and the beam is taken to be inextensible.
The beam's configuration is then described by the tangent angle $\theta(s)$, and the curvature is given by $\kappa(s) = \theta'(s)$, which characterizes the degree of bending. The equilibrium of the beam is formulated in terms of the internal moment and its balance with the distributed magnetic torque, leading to an integral representation of the governing equations.

The equilibrium of the beam follows from the balance of moments, consistent with the principle of virtual work~\citep{zhao2019mechanics}. As shown in Fig.~\ref{fig:1}(b), this moment balance can be expressed in integral form as
\begin{equation}
    \int_s^L m(x)\,\mathrm dx + M_\mathrm L - ((\boldsymbol{r}(L)-\boldsymbol{r}(s))\times \boldsymbol{F})\cdot \mathbf{e}_z
    = D(s)\kappa(s),
    \label{eq:1}
\end{equation}
where each term has a clear physical interpretation:
\begin{itemize}
    \item $m(s) = A B M \sin(\theta(s) -\psi(s) - \varphi)$ is the distributed magnetic torque per unit length, arising from the interaction between the remnant magnetization and the applied magnetic field, where $A$ is the cross-sectional area;
    \item $M_L$ is the moment applied at the boundary, $s=L$;
    \item $\boldsymbol{F} = (F_x, F_y)$ is the force applied at $s=L$, which contributes to the moment through the lever arm $(\boldsymbol{r}(L)-\boldsymbol{r}(s))$;
    \item $\mathbf{e}_x$ is the unit vector along horizontal direction, $\mathbf{e}_y$ is the unit vector along vertical direction and $\mathbf{e}_z=\mathbf{e}_x\times\mathbf{e}_y$;
    \item $\boldsymbol{r}(s) = \left( \int_0^s \cos\theta(u)\,\mathrm du,\ \int_0^s \sin\theta(u)\,\mathrm du \right)$ is the centerline position;
    \item $D(s) = E I(s)$ is the bending stiffness of the beam, in which $E$ is the Young's modulus (assumed constant) and $I(s)$ is the second moment of area. 
\end{itemize}

For a rectangular cross-section of width $w(s)$ and thickness $h(s)$, $I(s)=w(s)h(s)^3/12$. In general, varying the strip thickness is an efficient, but very challenging, means of changing $I(s)$. In practice, we shall consider strips with tapered widths, $w(s)$, but uniform thickness, $h(s)=\mathrm{cst}$.

\subsection{Dimensionless formulation}
\label{sec:Dimensionless_formulation}

The governing equations are cast in dimensionless form by scaling all lengths by the beam length $L$, i.e.~we let

\begin{equation}
\bar{s}=\frac{s}{L},\quad \bar{w}(\bar{s})=\frac{w(\bar{s})}{L},\quad \bar{h}=\frac{h}{L},\quad \bar{\kappa}(\bar{s})=\kappa (\bar{s})L.
\label{eq:2}
\end{equation}
Dimensionless variables are denoted by an overbar. Energies and moments are scaled by a characteristic elastic bending energy
\begin{equation}
P_0 = \frac{E L^3}{12}.
\label{eq:3}
\end{equation}
The corresponding characteristic force scale is $P_0/L$, obtained from the elastic energy scale, and is used to nondimensionalize forces. This yields the dimensionless quantities
\begin{equation}
\bar{F}_x=\frac{F_xL}{P_0},\quad \bar{F}_y=\frac{F_yL}{P_0},\quad \bar{M}_1=\frac{M_L}{P_0}.
\label{eq:4}
\end{equation}
By using these dimensionless variables, Eq.~\eqref{eq:1} transforms to
\begin{equation}
\int_{\bar{s}}^1{\alpha  \bar{w}(u)\sin \left( \theta (u) -\psi(u)-\varphi \right) \mathrm du}+\bar{F}_x\int_{\bar{s}}^1{\sin \theta (u) \mathrm du}-\bar{F}_y\int_{\bar{s}}^1{\cos \theta (u) \mathrm du}+\bar{M}_1=\beta \bar{w}(\bar{s})\bar{\kappa}(\bar{s}),
\label{eq:5}
\end{equation}
where 
\begin{equation}
    \alpha = {B M L^3 \bar{h}}/{P_0}=12BM/E\bar{h},
\end{equation}  is the fundamental magnetic coupling parameter and quantifies the ratio of typical magnetic torques to elastic restoring moments. We also introduce \begin{equation}
    \beta = \bar{h}^3,
\end{equation}
for convenience.

This dimensionless formulation shows that the  competition between magnetic actuation and elastic bending stiffness is characterized by $\alpha$: when $\alpha \ll 1$, the response is dominated by elastic effects and approaches that of a classical elastica; when $\alpha \gg 1$, magnetic actuation becomes dominant and significantly alters the deformation of the beam. 

To simplify the analysis, we first consider the case in which the magnetization direction is aligned with the local tangent of the elastica, so that $\psi(\bar{s})=0$. The generalization to spatially non-uniform magnetization directions is discussed in Section~\ref{sec:extension}.

\subsection{Clamped--free boundary condition}
\label{sec:Clamped_free_BC}

We  consider first the case where the beam is clamped at $\bar{s}=0$ and free at $\bar{s}=1$, with no external load applied at the free end (i.e., $\bar{F}_x=\bar{F}_y=0, \bar{M}_1=0$). This configuration represents a cantilever beam subject only to distributed magnetic loading, as arises in applications such as magnetic micro-actuators and sensors. Under these conditions, Eq.~\eqref{eq:5} simplifies considerably to
\begin{equation}
\beta \theta' (\bar{s}) \bar{w}(\bar{s})-\alpha \int_{\bar{s}}^1{\bar{w}(u)\sin\mathrm{(}\theta (u)}-\varphi )\mathrm du=0,
    \label{eq:6}
\end{equation}
where $\theta'(\bar{s})=\bar{\kappa}(\bar{s})$.

We assume that a desired target shape, $\theta(\bar{s})$, is known and treat \eqref{eq:6} as an equation for the unknown strip width $w(s)$ required to obtain this desired shape under the action of the imposed magnetic field. To render this equation more tractable, we differentiate with respect to \(\bar{s}\), yielding a first-order differential equation for the width function $\bar{w}(\bar{s})$
\begin{equation}
    \bar{w}'(\bar{s}) + \mathcal{A}(\bar{s}) \bar{w}(\bar{s}) = 0,
    \label{eq:7}
\end{equation}
where the coefficient function
\begin{equation}
    \mathcal{A}(\bar{s}) = \frac{\beta\theta''(\bar{s}) 
    + \alpha \sin(\varphi-\theta(\bar{s})))}
    {\beta \theta'(\bar{s})},
    \label{eq:8}
\end{equation}
encapsulates the combined influence of curvature variation and magnetic loading. This equation admits the general solution
\begin{equation}
    \bar{w}(\bar{s}) = w_1 \exp\left(\int_{\bar{s}}^1 \mathcal{A}(u)\mathrm du\right),
    \label{eq:9}
\end{equation}
with $w_1 = \bar{w}(1)$ denoting the beam's width at the free end. The expression in \eqref{eq:9} reveals that the width variation along the beam required to realize a desired shape is very (exponentially) sensitive to the integrated effect of the mechanical-magnetic coupling.

It is worth emphasizing that many existing inverse-design theories for elastica are derived from the differential form of the governing equations~\citep{liu2020tapered,fan2020inverse,zhang2022shape}. While such formulations correctly describe local equilibrium by relating curvature, loading, and structural properties pointwise, they do not by themselves expose the global constraints imposed by boundary conditions. Consequently, a locally obtained width distribution does not necessarily guarantee that the prescribed target shape is physically attainable.

This limitation can be illustrated by considering the inverse design of a circular arc with constant nonzero curvature, i.e., $\theta'(\bar{s})=\mathrm{const}\neq 0$. In a differential-form-based framework, substituting this target curvature into the local inverse-design equation \eqref{eq:9} yields a corresponding width distribution, and the target shape may therefore appear to be admissible. However, this conclusion is physically incorrect for a clamped--free magneto-elastica with finite bending stiffness: the free end is moment-free, and hence the curvature must vanish at the tip. A circular arc with nonzero constant curvature violates this boundary condition and cannot be realized as an equilibrium configuration, for any width distribution. This example shows that local consistency alone is insufficient for inverse design. In contrast, the integral formulation developed in this work explicitly incorporates the boundary-induced global constraints, thereby excluding such nonphysical solutions and enabling a direct characterization of shape attainability and the admissible design space for the magneto-elastica.

To determine the design space of a magneto-elastica subject to clamped--free boundary conditions, we next derive the constraints that the target rotation field $\theta(\bar{s})$ must satisfy. First, it follows from Eq.~\eqref{eq:6} that $\theta'(\bar{s})=0$ at $\bar{s}=1$. Second, Eq.~\eqref{eq:6} also implies that $\theta'(\bar{s})>0$ when $\theta(\bar{s})<\varphi$. Finally, because the free-end condition requires $\theta'(1)=0$, Eq.~\eqref{eq:8} would become singular unless
$\beta\theta''(1)+\alpha\sin(\varphi-\theta(1))=0$. In addition, the clamped condition at the left end requires the initial rotation to satisfy $\theta(0)=0$.

We summarize the constraints for $\theta(\bar{s})$ as follows
\begin{equation}
\begin{cases}
    \beta\theta''(1) + \alpha \sin(\varphi-\theta(1)) = 0, \\
    \theta'(1) = 0, \\
    \theta(\bar{s})<\varphi, \\
    \theta'(\bar{s}) > 0, \\
    \theta(0) = 0.
\end{cases}
\label{eq:10}
\end{equation}

In principle, $\theta(\bar{s})$ can be any smooth function satisfying these constraints, leading to an infinite-dimensional inverse-design problem. Rather than approximating the desired shape numerically, we instead introduce a family of shapes with four degrees of freedom; specifically we introduce a cubic polynomial representation as a low-dimensional parametrization of the target rotation field. This reduces the search space to a finite set of design parameters while preserving the dominant geometric features of the desired deformation. We assume that $\theta(\bar{s})$ takes the form
\begin{equation}
    \theta(\bar{s}) = a + b(\bar{s}-1) + c(\bar{s}-1)^2 + d(\bar{s}-1)^3,
    \label{eq:11}
\end{equation}
where $a=\theta(1)$ is the rotation angle at the free end. This kinematic assumption allows the essential boundary conditions to be satisfied while retaining sufficient flexibility to capture the target deformation shape. Substitution into the constraints in Eq.~\eqref{eq:10} gives the following conditions
\begin{equation}
\begin{cases}
    a < \varphi, \\
    b = 0, \\
    c = \dfrac{\alpha}{2}\sin(a-\varphi), \\
    d = a + \dfrac{\alpha}{2}\sin(a-\varphi), \\
    a + \dfrac{\alpha}{6}\sin(a-\varphi) > 0.
\end{cases}
\label{eq:12}
\end{equation}
These relations reveal that the deformation is primarily governed by the free-end angle $\theta(1)$, which itself depends on the magnetic coupling parameter $\alpha$ and the field direction $\varphi$. The inequality constraints ensure physical admissibility of the solution. 

\begin{figure}[h]
    \centering
    \includegraphics[width=\textwidth]{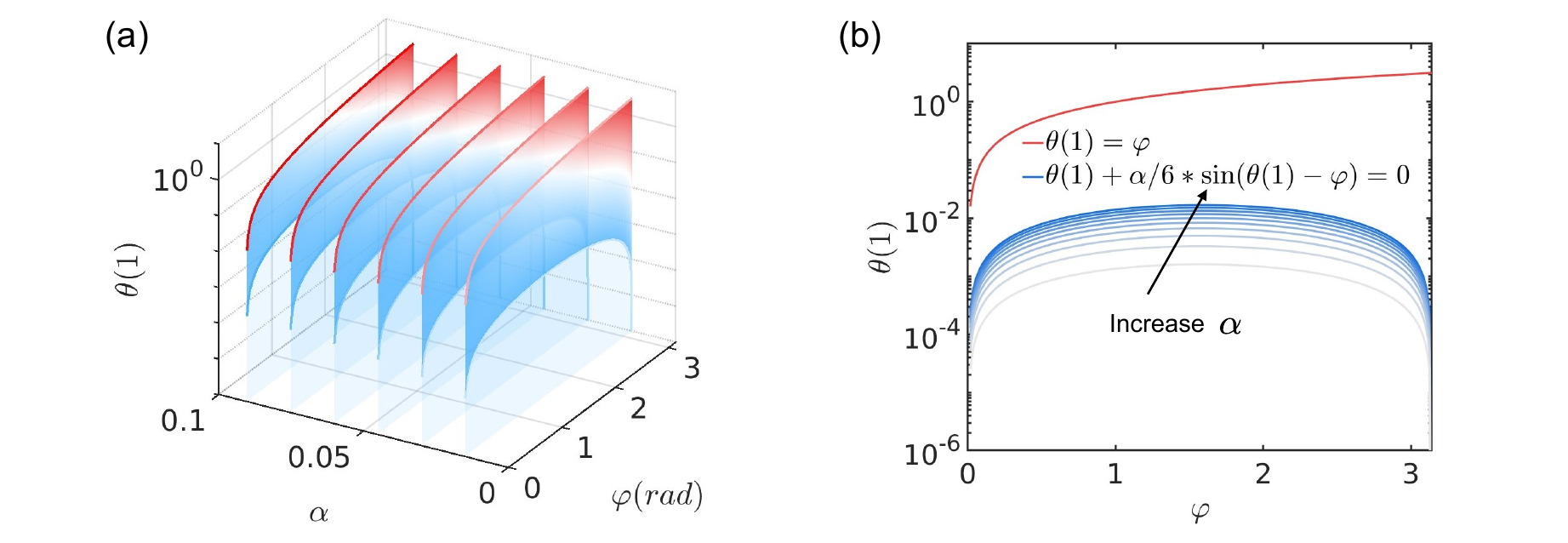}
    \caption{\textbf{Admissible design space of a magneto-elastica with clamped--free boundary conditions.} 
    (a) Admissible region of $\theta(1)$ as a function of $\alpha$ and $\varphi$. For given values of $\alpha$ and $\varphi$, admissible solutions are confined to the shaded region bounded by the two constraint surfaces. The upper boundary (red) corresponds to $\theta(1)=\varphi$, while the lower boundary (blue) is defined by $\theta(1)+\alpha \sin(\theta(1)-\varphi)/6=0$, and hence depends on $\alpha$. 
    (b) Projection of the admissible design space onto the $(\theta(1), \varphi)$ plane for different values of $\alpha$. As $\alpha$ increases, the admissible region shrinks, indicating a reduction in the feasible inverse-design space.}
    \label{fig:2}
\end{figure}

The admissible region defined by Eq.~\eqref{eq:12} can be visualized in the $(\theta(1), \varphi, \alpha)$ space, as shown in Fig.~\ref{fig:2}(a). For given values of $\alpha$ and $\varphi$, feasible inverse designs correspond to values of $\theta(1)$ lying within a bounded region defined by two constraint surfaces. The upper boundary (red) corresponds to $\theta(1)=\varphi$, while the lower boundary (blue) satisfies $\theta(1) + \frac{\alpha}{6}\sin(\theta(1)-\varphi)=0$.
Figure~\ref{fig:2}(b) shows the projection of this admissible region onto the $(\theta(1), \varphi)$ plane for different values of $\alpha$. As $\alpha$ increases, the lower boundary shifts upward, leading to a progressive reduction of the admissible design space. This trend indicates that stronger magnetic actuation imposes more restrictive constraints on feasible inverse designs, ultimately limiting the range of achievable target shapes.

\subsection{Clamped--clamped boundary condition}
\label{sec:Clamped_clamped_BC}

We now analyze the case in which both ends are clamped. This configuration introduces additional constraints that substantially alter the mechanical behavior, making it more complex than the clamped--free case considered above. 
We retain the inextensibility assumption for the clamped--clamped configuration, owing to the slender geometry of the elastica and the moderate magnetic-field strength considered here, which is comparable to that used in previous work~\citep{zhang2025snap}. This assumption has been validated for similar magneto-elastic systems~\citep{abbasi2023snap,zhang2025snap}, and a detailed scaling analysis is provided in \ref{app:C}.
In contrast to the clamped--free case, where the free end can rotate freely, both ends are now geometrically constrained. Reaction forces $F_x$ and $F_y$ are therefore generated at the supports to maintain the imposed boundary conditions. These reaction forces significantly complicate the mechanics, as they introduce additional moments that vary along the beam length. We therefore differentiate the full form of the elastica equation, Eq.~\eqref{eq:5}, to obtain the equation for the width profile $w(s)$ required to generate the target rotation field $\theta(s)$. Following this procedure, the governing equation for $w(s)$ takes the form
\begin{equation}
    \bar{w}'(\bar{s}) + \mathcal{A}(\bar{s})\bar{w}(\bar{s}) + \mathcal{B}(\bar{s}) = 0,
    \label{eq:13}
\end{equation}
where 
\begin{equation}
\begin{aligned}
    \mathcal{B}(\bar{s}) &= \frac{\bar{F}_x \sin\theta(\bar{s}) - \bar{F}_y \cos\theta(\bar{s})}
    {\beta\theta'(\bar{s})}.
\end{aligned}
\label{eq:14}
\end{equation}
The function $\mathcal{A}(\bar{s})$ is as before, see \eqref{eq:8}, captures the same mechanical-magnetic coupling as in the clamped--free case in Eq.~\eqref{eq:8}, while $\mathcal{B}(\bar{s})$ incorporates the additional effect of the reaction forces.

The linear ordinary differential equation for \(\bar{w}(\bar{s})\), \eqref{eq:14}, may readily be solved using an integrating factor $1/p(\bar{s})$ where $p(\bar{s}) = \exp\left(-\int_0^{\bar{s}} \mathcal{A}(u)\mathrm du\right)$; we may write the general solution as
\begin{equation}
    \bar{w}(\bar{s}) = p(\bar{s})\left( \mathcal{C} - \int_0^{\bar{s}} \frac{\mathcal{B}(u)}{p(u)}\mathrm du \right),
    \label{eq:15}
\end{equation}
where $\mathcal{C}$ is an integration constant. This expression illustrates how the width variation results from the competing influences of the magneto-elastic coupling (through $p(\bar{s})$) and the reaction forces (through the integral term).

To determine the unknown reactions $\bar{F}_x$ and $\bar{F}_y$, we must impose appropriate BCs; moreover, since there are two additional unknowns, we must impose two further conditions. Unlike the clamped--free case, where the width is imposed through a single boundary condition (at the free end), we now require compatibility at both ends and an intermediate point. Specifically, we impose $\bar{w}(0) = w_0$, $\bar{w}(1) = w_1$, and an intermediate condition $\bar{w}(\gamma) = w_\gamma$ for some \(\gamma \in (0,1)\).
Defining the auxiliary functions
\begin{equation}
\begin{aligned}
    D_1(\bar{s}) &= \int_0^{\bar{s}}  \frac{\sin\theta(u)}{\theta'(u) p(u)}\mathrm du, \\
    D_2(\bar{s}) &= -\int_0^{\bar{s}}  \frac{\cos\theta(u)}{\theta'(u) p(u)}\mathrm du,
\end{aligned}
\label{eq:16}
\end{equation}
which physically represent weighted integrals of the trigonometric functions that modulate the force components, we obtain explicit expressions for the reaction forces
\begin{equation}
\begin{cases}
    \mathcal{C} = w_0, \\
    F_y = \dfrac{ (w_0 - w_1/p(1))D_1(\gamma) - D_1(1)(w_0 - w_\gamma/p(\gamma)) }
                { D_2(1)D_1(\gamma) - D_1(1)D_2(\gamma) }, \\
    F_x = \dfrac{ (w_0 - w_1/p(1))D_2(\gamma) - D_2(1)(w_0 - w_\gamma/p(\gamma)) }
                { D_1(1)D_2(\gamma) - D_2(1)D_1(\gamma) }.
\end{cases}
\label{eq:17}
\end{equation}
These relations complete the mathematical description of the clamped--clamped case, revealing how the reaction forces depend on the width BCs, the deformation geometry through $\theta(\bar{s})$, and the magnetic-elastic coupling through $p(\bar{s})$.

\subsection{Reduction to classical inverse design of elastica}
\label{sec:Reduction_to_classical}

In this section, we briefly demonstrate that the above framework reduces to the classical inverse design theory of an elastica in the absence of a magnetic field, i.e.~when $\alpha=0$. Since there is no magnetic field, we consider the clamped--clamped case of a naturally straight beam. Under these conditions, Eq.~\eqref{eq:5} reduces to
\begin{equation}
\beta \,\bar{w}(\bar{s})\,\theta ' (\bar{s})-\bar{F}_x\int_{\bar{s}}^1{\sin \theta (u)\,\mathrm du}+\bar{F}_y\int_{\bar{s}}^1{\cos \theta (u)\,\mathrm du}-\bar{M}_1=0,
\label{eq:19}
\end{equation}
where $\bar{x}' = \cos \theta$ and $\bar{y}' = \sin \theta$. Using these relations, Eq.~\eqref{eq:19} can be rewritten as
\begin{equation}
\bar{w}(\bar{s})=\frac{\bar{M}_1-\bar{F}_y\left( x_1-\bar{x}(\bar{s}) \right)+\bar{F}_x\left( y_1-\bar{y}(\bar{s}) \right)}{\beta \,\theta' (\bar{s})},
\label{eq:20}
\end{equation}
with $x_1 = \bar{x}(1)$ and $y_1 =\bar{y}(1)$ denoting the coordinates of the beam’s right end.  
In the special case $\bar{F}_x = 0$, $w(\bar{s})$ can be determined analytically by imposing the BCs $\bar{w}(0) = w_0$ and $\bar{w}(1) = w_1$, which gives
\begin{equation}
\bar{w}(\bar{s})=\frac{\theta ' (1)\,w_1\,\bar{y}(\bar{s})+(y_1-\bar{y}(\bar{s}))\,w_0\,\theta' (0)}{y_1\,\theta' (\bar{s})}.
\label{eq:21}
\end{equation}
Here, we have chosen the left end of the beam as the origin, i.e., $y(0) = 0$. Eq.~\eqref{eq:21} is equivalent to Eq.~(14) of \cite{liu2020tapered}, confirming that our framework consistently reduces to the inverse design theory of a classical elastica studied previously \citep{liu2020tapered,fan2020inverse}.

\subsection{Tessellation}
\label{sec:Tessellation}

For an axially symmetric surface, an alternative design principle is to achieve a 3D surface in which $N$ strips (petals)  tessellate upon actuation and approximate the desired axially symmetric surface. As we shall see, the requirement to tessellate  constrains the width profile of the sheet, and so it is not possible to construct a desired surface that also tessellates without another variable, such as the thickness, being varied~\citep{liu2020tapered}. To provide this additional flexibility, we adopt a method to tailor the effective bending stiffness of the magneto-elastica by introducing holes along the centerline~\citep{zhang2022shape}. Under this framework, tessellation with a desired 3D shape can be realized by prescribing an appropriate effective width distribution while simultaneously ensuring conformal fitting of each individual petal. To achieve tessellation, an additional restriction on the width $\bar{w}(\bar{s})$ is introduced
\begin{equation}
       \bar{w}(\bar{s})=\bar{x}(\bar{s})\tan\frac{\pi}{N}.
    \label{eq:24}
\end{equation} 
This ensures that each of the $N$ petals touches its neighbour. After cutting holes along the centerline of the beam, the remaining portion is defined as the effective width $\bar{w}^e(\bar{s})$, and $\bar{w}^e(\bar{s})$ can be solved from Eq.~\eqref{eq:9}. Assume the width distribution of the holes is $\bar{w}_{\circ}(\bar{s})$, the effective width is
\begin{equation}
    \bar{w}^e(\bar{s})=\bar{w}(\bar{s})-\bar{w}_{\circ}(\bar{s}).
    \label{eq:25}
\end{equation}

The distribution of porosity $\lambda$ can be calculated as
\begin{equation}
    \lambda=\frac{\bar{w}_{\circ}(\bar{s})}{\bar{w}(\bar{s})}=
    1-\bar{w}^e(\bar{s})/\bar{w}(\bar{s}).
\label{eq:26}
\end{equation} Note that the width distribution of the holes can readily be determined as $\bar{w}_\circ(\bar{s})=\bar{w}(\bar{s})-\bar{w}^e(\bar{s})$ but the requirement that $\bar{w}_{\circ}(\bar{s})>0$ everywhere introduces some constraints on the choice of $w_1$.

Having established the general theory in a number of scenarios of interest, we now proceed to demonstrate it in particular cases.

\section{Illustrative examples and discussion}
\label{sec:res_dis}

In this section, we illustrate the general theoretical framework presented in \S2 through representative  examples of the inverse-design strategies derived in \S2. The performance of the designed width (and porosity) profiles is evaluated by comparing the resulting 3D shapes under magnetic actuation with the prescribed target configurations. Validation is carried out using a combination of physical experiments (see \ref{app:A} for more details) and numerical simulations based on the discrete elastic rod (DER) framework \citep{jawed2018primer,huang2025tutorial} (see \ref{app:B}). 
Results are presented for different boundary conditions and design criteria. To examine the role of magnetic actuation, we compare configurations obtained with and without an applied magnetic field, and quantify the deviation from the target shape through curvature-based metrics. This analysis further leads to a scaling relation for the deviation between the desired and obtained curvatures that result from mismatch in the magnetic field.

Unless otherwise specified, the following physical parameters are used: beam length $L \sim 0.04\,\mathrm{m}$, magnetization $M \sim 10^{5}\,\mathrm{A/m}$, magnetic field $B \sim 0.08\,\mathrm{T}$, Young’s modulus $E \sim 10^{6}\,\mathrm{Pa}$, and thickness $h \sim 0.002\,\mathrm{m}$. These correspond to dimensionless parameters $\beta \approx 1.25\times10^{-4}$. 
For different cases, $\beta$ and $\varphi = \pi/2$ are kept fixed, while $\alpha$ is varied to explore different regimes of magnetic-to-elastic coupling. In particular, $\alpha = 3 \times 10^{-4}$ is used for the clamped–free case, whereas $\alpha = 5 \times 10^{-3}$ is used for the clamped–clamped and tessellation cases.

\subsection{Clamped--free boundary condition}

\begin{figure}[!b]
    \centering
    \includegraphics[width=\textwidth]{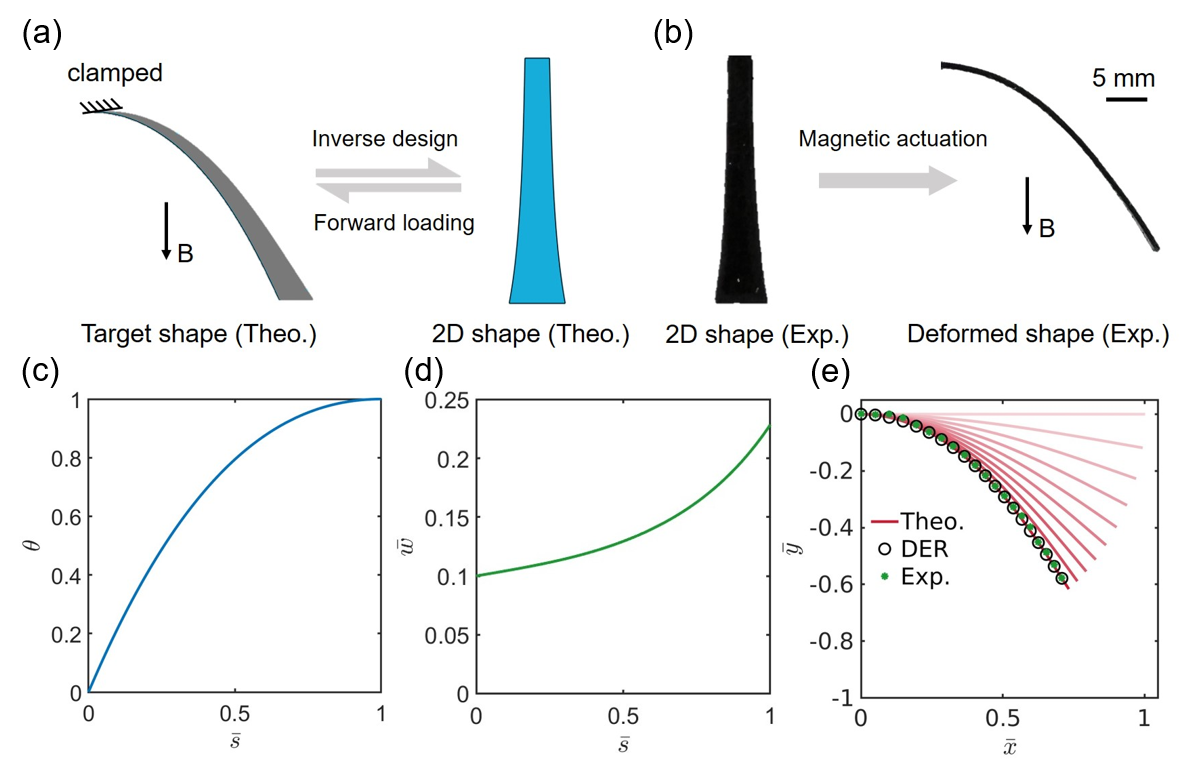}
    \caption{\textbf{Inverse design of a morphing magneto-elastica under clamped--free boundary conditions.}
    (a) The target shape (left) and the designed 2D tapering pattern predicted by theory (right). 
    (b) The experimental verification of clamped--free case. 
    (c) The rotation angle of the target shape. (d) The theoretical width distribution. 
    (e) The initial state and the actuated state of the designed beam. The final shape is verified by DER (black circle) and experiments (green star).}
    \label{fig:3}
\end{figure}

We first consider the clamped--free boundary condition and the magnetization is along the tangent direction ($\psi(\bar{s})=0$). The prescribed rotation profile $\theta(s)$, representing the target shape, must satisfy the admissibility constraints given in Eq.~\eqref{eq:12}. Setting $\theta(1) = 1$, we define a target profile as
\begin{equation}
\theta(\bar{s}) = 1 + \frac{\alpha}{2}\sin\left(1-\frac{\pi}{2}\right)(\bar{s}-1)^2 + \left(1 + \frac{\alpha}{2}\sin(1-\frac{\pi}{2})\right)(\bar{s}-1)^3.
\label{eq:29}
\end{equation}
The corresponding rotation profile is shown in Fig.~\ref{fig:3}(c), together with the resulting target shape in Fig.~\ref{fig:3}(a). Based on the inverse design framework, the associated width distribution $\bar{w}(\bar{s})$ is obtained, as shown in Fig.~\ref{fig:3}(d).

To validate the proposed framework, a magneto-elastica with the designed width distribution $\bar{w}(\bar{s})$ (Fig.~\ref{fig:3}(d)) is fabricated (Fig.~\ref{fig:3}(b)) and subjected to a uniform magnetic field. 
The vanishing curvature at the free end ($\kappa = 0$) leads to a monotonic increase in the beam width, compensating for the magnetically induced deformation. 
The deformed configurations obtained from theory, DER simulations, and experiments (Fig.~\ref{fig:3}(e)) are in excellent agreement with the desired target shape.
The chosen target profile lies within the admissible design space identified in Fig.~\ref{fig:2}, ensuring that a feasible inverse design exists.

\subsection{Clamped--clamped boundary condition}

\begin{figure}[!h]
    \centering
    \includegraphics[width=\textwidth]{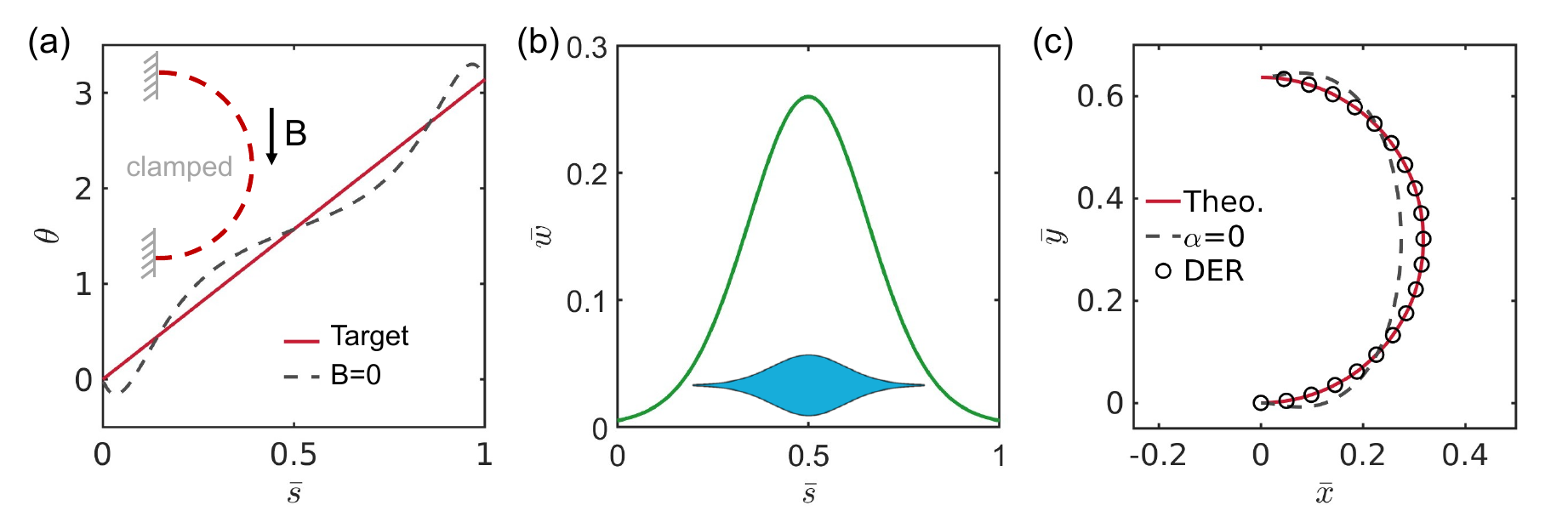}
    \caption{\textbf{Inverse design of a semi-circular arc under clamped--clamped boundary conditions.}
    (a) Prescribed rotation profile $\theta(\bar{s})$ corresponding to the target shape (inset). The solid curve denotes to the actuated case, while the dashed curve corresponds the unactuated configuration ($B=0$). 
    (b) Designed width distribution $\bar{w}(\bar{s})$ obtained from the inverse formulation. 
    (c) Comparison of the deformed configuration with theoretical prediction and DER simulations. The red curve denotes the target shape, the dashed curve corresponds to the unactuated configuration, and the symbols represent DER results.}
    \label{fig:4}
\end{figure}

\begin{figure}[h]
    \centering
    \includegraphics[width=\textwidth]{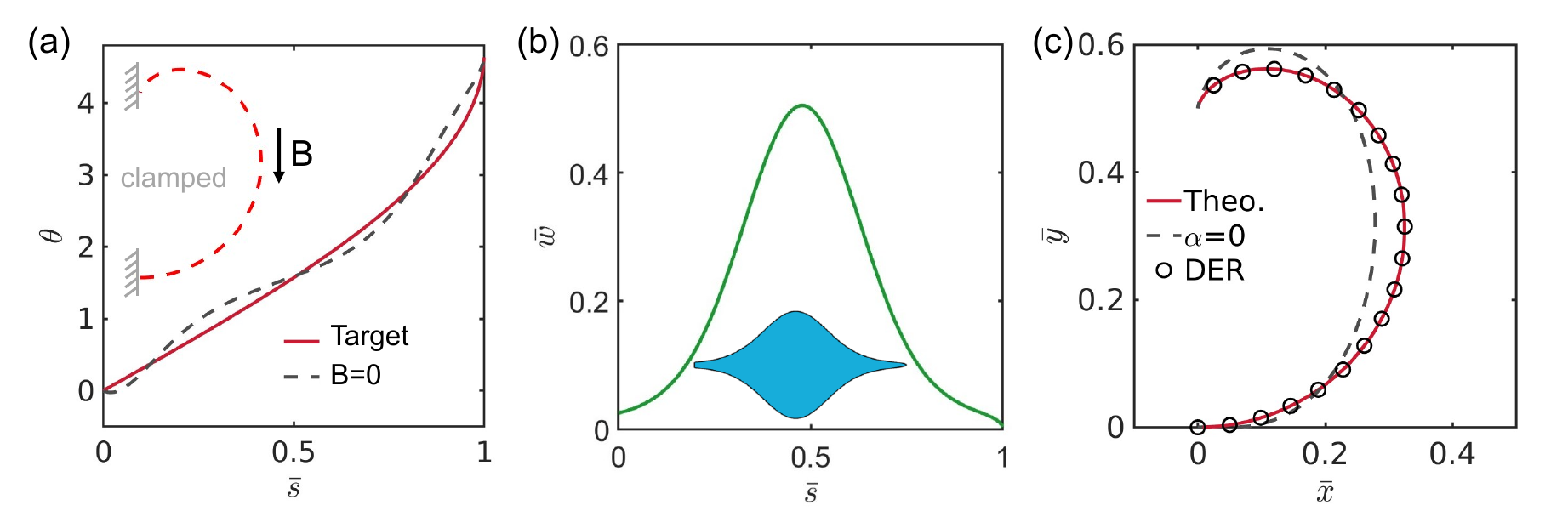}
    \caption{\textbf{Inverse design of a cardioid shape under clamped--clamped boundary conditions.} 
    (a) 
    Prescribed rotation profile $\theta(\bar{s})$ corresponding to the target shape (inset). The solid curve denotes to the actuated case, while the dashed curve corresponds the unactuated configuration ($B=0$). 
    (b) Designed width distribution $\bar{w}(\bar{s})$ obtained from the inverse formulation. 
    (c) Comparison of the deformed configuration with theoretical prediction and DER simulations. The red curve denotes the target shape, the dashed curve corresponds to the unactuated configuration, and the symbols represent DER results.}
    \label{fig:5}
\end{figure}

For clamped--clamped boundary conditions, a greater range of target shapes can be achieved due to the additional unknowns introduced at the second clamped boundary (the reaction forces and moment). To illustrate this increased flexibility, we consider two representative target shapes: a semi-circular arc (Fig.~\ref{fig:4}) and a cardioid (Fig.~\ref{fig:5}).

For the semi-circular arc, the prescribed rotation profile is $\theta(\bar{s})=\pi\bar{s}$, as shown in Fig.~\ref{fig:4}(a). Using Eq.~\eqref{eq:15} with parameters $w_0 = w_1 = 0.005$, $\gamma=0.5$, and $w_{\gamma} = 0.26$ to ensure the obtained $\bar{w}(\bar{s})$ is larger than zero, we observe the resulting width distribution (Fig.~\ref{fig:4}(b)) is larger at the midpoint than near the ends. The corresponding deformed configuration, obtained from DER simulations (Fig.~\ref{fig:4}(c)), shows excellent agreement with the prescribed target shape. 

In contrast to the clamped--free case, the unactuated configuration ($\alpha=0$) is already deformed by the clamped boundary conditions. The dashed curve in Fig.~\ref{fig:4}(c) shows this unactuated shape, indicating that magnetic actuation transforms the clamped–clamped buckling mode into the desired semi-circular arc.

For the cardioid target, the rotation profile $\theta(\bar{s})$ is computed from the parametric representation~\citep{abbena2017modern,li2025biomimetic}
\begin{equation}
    \begin{cases}
        x=0.25(1-\cos(\pi(1-t)))\sin(\pi(1-t)),&\\
        y=0.25(1-\cos(\pi(1-t)))\cos(\pi(1-t)),
    \end{cases} t\in(0,1)
    \label{eq:30}
\end{equation}
via $\theta(\bar{s})=\arctan\!\big(y'(t)/x'(t)\big)$, as shown in Fig.~\ref{fig:5}(a). The parameters in Eq.~\eqref{eq:15} are set as $w_0 = 0.025$, $w_1 = 0.0035$, $\gamma=0.5$, and $w_{\gamma} = 0.5$, leading to the width profile shown in Fig.~\ref{fig:5}(b). 

The resulting deformation, validated by DER simulations (Fig.~\ref{fig:5}(c)), agrees well with the prescribed cardioid shape. As in the semi-circular case, the dashed curve represents the unactuated configuration, showing that magnetic actuation guides the structure from the clamped–clamped buckled state toward the target geometry.

These examples demonstrate that the proposed framework can accommodate a broader class of admissible shapes under clamped--clamped boundary conditions.

\subsection{Errors introduced by the magnetic field deviation}

The effect of magnetic actuation is illustrated in Figs~\ref{fig:4} and  \ref{fig:5}. To further quantify the sensitivity of the final shape to variations in the applied magnetic field, we introduce a scaling law relating curvature deviation to changes in the dimensionless parameter $\alpha$. 

\begin{figure}[h]
    \centering
    \includegraphics[width=\textwidth]{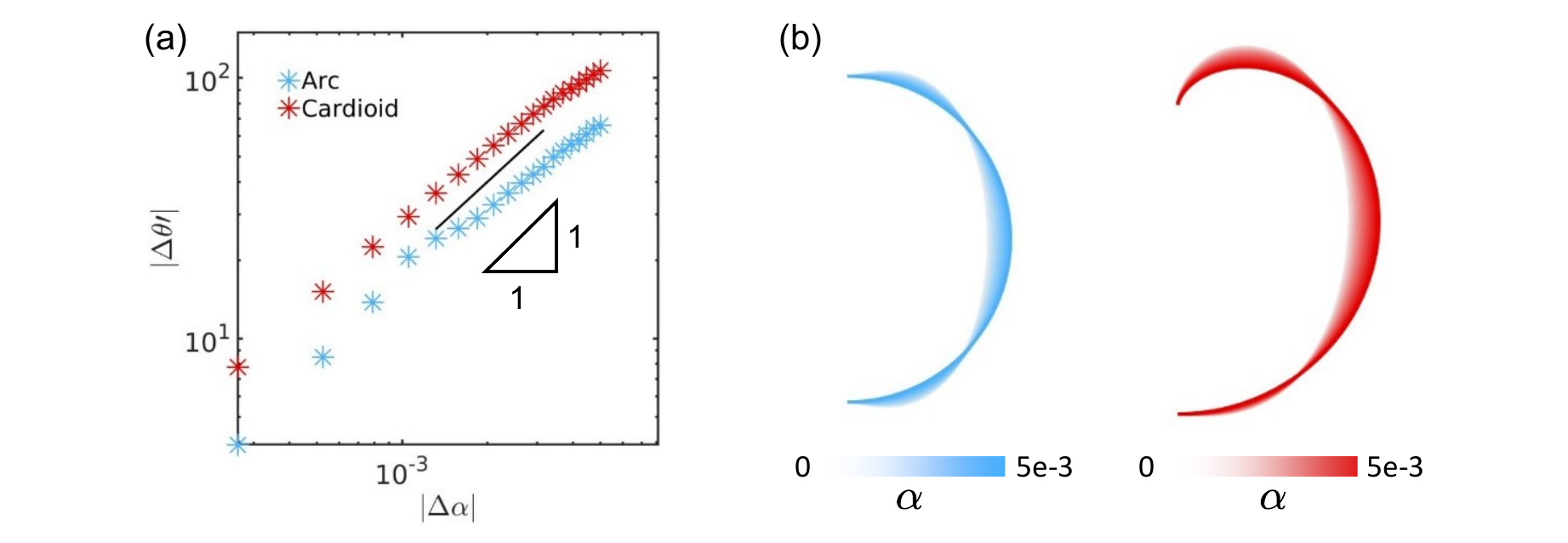}
    \caption{\textbf{Scaling of curvature error with respect to magnetic field mismatch.} 
    (a) Relationship between the maximal curvature deviation $|\Delta \theta'|$ and the variation in magnetic field parameter $\Delta \alpha$. The results exhibit a linear scaling, $|\Delta \theta'| \propto |\Delta \alpha|$, for both the arc and cardioid cases. 
    (b) Deformed configurations of magneto-elastica for different values of $\alpha$, illustrating the sensitivity of the shape to variations in magnetic actuation.}
    \label{fig:6}
\end{figure}

The deformation of the magneto-elastica is governed by the balance between magnetic and elastic moments. A perturbation in the magnetic field modifies the effective magnetic torque, leading to a deviation in curvature from the target configuration. To quantify this deviation, we use the $L^2$ norm of the curvature difference, i.e.~we define
\begin{equation}
    |\Delta\theta'| = \sqrt{\int_0^1 \big(\theta'(\bar{s}) - \theta_0'(\bar{s})\big)^2 \,\mathrm{d}\bar{s}},
    \label{eq:31}
\end{equation}
where $\theta'(\bar{s})$ denotes the curvature under the applied magnetic field, and $\theta_0'(\bar{s})$ the target curvature.
A linear scaling relationship is observed from the moment equilibrium between magnetic moment, $ABM \sin(\varphi-\theta)\,\Delta L$ and elastic moment, $D \theta'(s)$, where $\Delta L$ denotes a small segment of the micro-element length:
\begin{equation}
    |\Delta\theta'(\bar{s})| \sim |\Delta\alpha|,
    \label{eq:32}
\end{equation}
where $\Delta\alpha$ denotes the deviation of the dimensionless parameter $\alpha$ from its reference value. 

This scaling law is validated against numerical results in Fig.~\ref{fig:6}, showing excellent agreement. Physically, this indicates that the curvature deviation exhibits a first-order linear dependence on variations in magnetic actuation.

\subsection{Tessellation}

\begin{figure}[!h]
    \centering
    \includegraphics[width=\textwidth]{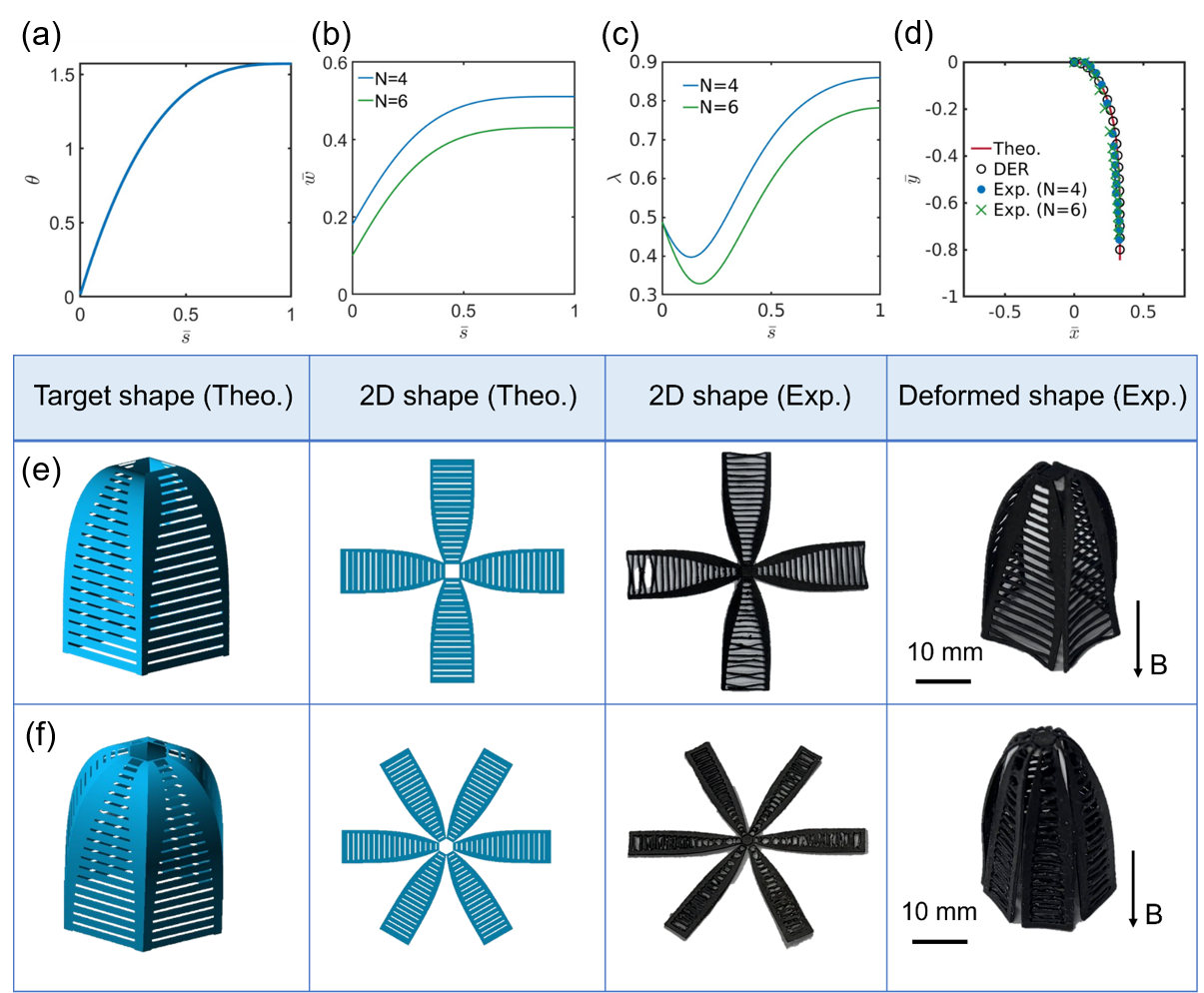}
    \caption{\textbf{Inverse design of tessellated morphing magneto-elastica under clamped--free boundary conditions.} 
    (a) Prescribed rotation profile $\theta(\bar{s})$ of the target shape. 
    (b) Width distributions $\bar{w}(\bar{s})$ predicted for $N=4$ and $N=6$ petals with $\alpha=5\times10^{-4}$.
    (c) Corresponding porosity distributions required to realize the target shape. 
    (d) Comparison of the deformed configurations from theory, DER simulations, and experiments, showing good agreement. 
    (e,f) Realization of tessellated morphing structures for (e) $N=4$ and (f) $N=6$: target 3D shape (theory), designed 2D pattern, fabricated sample, and resulting deformed configuration under magnetic actuation. The beams are clamped in the center when actuated in magnetic field.}
    \label{fig:7}
\end{figure}

To extend the proposed inverse design framework beyond single slender structures, we consider the tessellation of magneto-elastica, where multiple units are assembled to form discretized morphing surfaces with prescribed rotational symmetry. 

We consider tessellated structures under clamped--free boundary conditions. The parameters are chosen as $d = a = \varphi$, $\varphi = \pi/2$, $b=c=0$, and $\alpha=5\times10^{-4}$, such that the target rotation profile in Eq.~\eqref{eq:12} reduces to
\begin{equation}
    \theta(\bar{s}) = \frac{\pi}{2} + \frac{\pi}{2} (\bar{s}-1)^3,
    \label{eq:33}
\end{equation}
as shown in Fig.~\ref{fig:7}(a). 

For this target shape, we consider two tessellation configurations with different rotational symmetries, corresponding to $N=4$ and $N=6$ petals. Based on these symmetries, the corresponding width distributions are designed (Fig.~\ref{fig:7}(b)), and the associated porosity distributions are obtained using Eq.~\eqref{eq:26} (Fig.~\ref{fig:7}(c)). 
Using Eqs.~\eqref{eq:24} and \eqref{eq:25}, the resulting 3D geometries are constructed for both cases, as shown in Fig.~\ref{fig:7}(e,f). The deformed configurations obtained from theory, DER simulations, and experiments are compared in Fig.~\ref{fig:7}(d), demonstrating good agreement with the prescribed target shape. 

We note that, due to limitations in the 3D printing process, the central hole in the design is filled in the experimental realizations. This modification introduces small deviations from the theoretical prediction, consistent with observations reported by \citet{barckicke2025localization}. 

These results demonstrate that the proposed framework can be extended from individual slender structures to discretized morphing surfaces with prescribed symmetry, highlighting its potential for the design of programmable architected materials and soft robotic systems.

\section{Non-uniform magnetization direction}
\label{sec:extension}

In this section, we revisit the simplifying assumption that $\psi(s)$, i.e.~that the magnetization be uniform and aligned parallel to the tangent direction. We will show that the proposed inverse-design framework can be extended straightforwardly to the magneto-elastica with a spatially prescribed magnetization direction. Note that this extension is applicable only to hard magnetorheological elastomers (MREs) and cannot be extended to soft MREs, because the prescribed magnetization in soft MREs is altered by the applied magnetic field during actuation~\citep{wu2026review}.

We assume that the inclination angle of the magnetization, $\psi(\bar{s})$, is prescribed. The coefficient $\mathcal{A}(\bar{s})$ in Eq.~\eqref{eq:8} is then modified by replacing $\theta$ with $\theta-\psi$ in the magnetic torque term, yielding
\begin{equation}
\mathcal{A}(\bar{s}) =
\frac{\beta\theta''(\bar{s})
+\alpha\sin\left(\varphi-\theta(\bar{s})+\psi(\bar{s})\right)}
{\beta\theta'(\bar{s})}.
\label{eq:34}
\end{equation} Eq. \eqref{eq:9} then gives the width profile required to give the desired 3D shape.

\begin{figure}[!h]
    \centering
    \includegraphics[width=\textwidth]{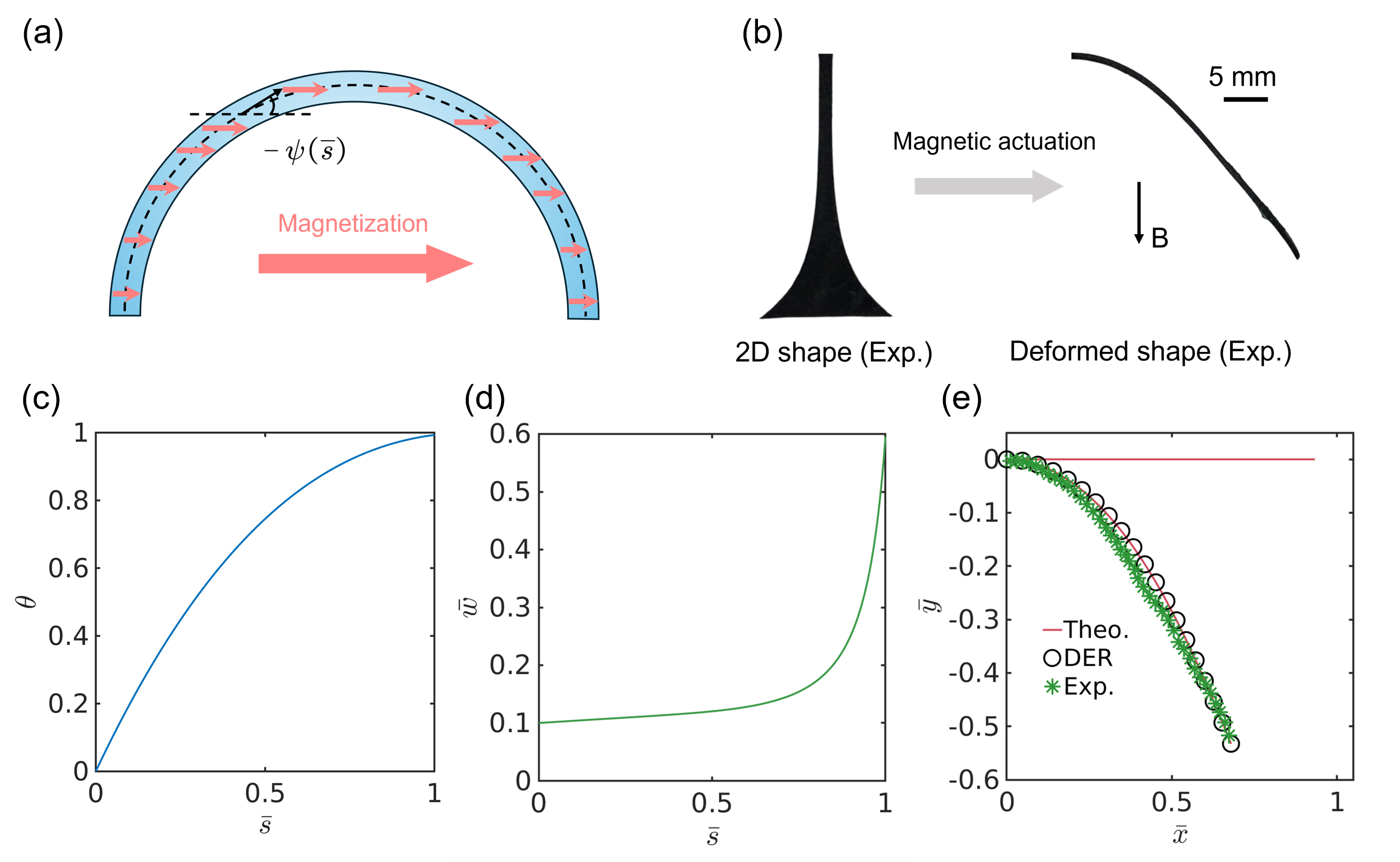}
    \caption{\textbf{Inverse design of a morphing magneto-elastica with non-uniform magnetization vector under clamped--free boundary conditions.}
    (a) A schematic illustration of how  a spatially varying magnetization profile is obtained: the strip is bent into a desired shape and then a magnetic field applied. 
    (b) The experimental verification of clamped--free case. 
    (c) The rotation angle of the target shape. (d) The theoretical width distribution. 
    (e) The initial state and the actuated state of the designed beam. The final shape is verified by DER (black circles) and experiments (green star).}
    \label{fig:rev}
\end{figure}

As an illustrative example, we consider a straight beam that is magnetized by first bending it around a circular mould to form a semicircle and then applying a strong field in the horizontal direction; this results in the magnetization profile $\psi(\bar{s})=\pi(\bar{s}-0.5)$, as illustrated in Fig.~\ref{fig:rev}(a). Here, similar to the case in Section~\ref{sec:Clamped_free_BC}, we choose a cubic function satisfying $\theta(0)=\theta'(1)=0$ as the target configuration: $\theta(\bar{s})=1-0.648(\bar{s}-1)^2+0.352(\bar{s}-1)^3$,  as shown in Fig.~\ref{fig:rev}(c). Substituting the prescribed magnetization profile into Eqs.~\eqref{eq:34} and \eqref{eq:11} yields the corresponding width distribution, shown in Fig.~\ref{fig:rev}(d). Based on this design, we fabricate the magneto-elastica, as shown in Fig.~\ref{fig:rev}(b). The resulting equilibrium configurations obtained from both DER simulations and experiments agree well with the theoretical prediction, as shown in Fig.~\ref{fig:rev}(e). 

Note that extending the design framework to non-uniform magnetization enables target shapes beyond the admissible design space of uniformly magnetized magneto-elastica, thereby offering considerable potential for diverse applications.

\section{Conclusion}
\label{conclusion}

In this work, we have developed an analytical inverse-design framework for the magneto-elastica based on the integral form of the moment equilibrium equations. Compared with approaches based purely on the differential-form~\citep{lum2016shape,liu2020tapered,fan2020inverse}, the proposed formulation explicitly incorporates the global constraints that are imposed by boundary conditions and magnetic actuation. This allows the framework to distinguish physically attainable target shapes from locally consistent, but globally unattainable, solutions. As a result, it enables a systematic characterization of the admissible design space and provides explicit attainability criteria for clamped--free configurations.

We have validated our framework  through experiments and DER simulations on representative semi-circular and cardioid target shapes; error-scaling analysis confirms the quantitative robustness of this approach. We further demonstrate its versatility by extending it to tessellated morphing surfaces with prescribed symmetries using a combination of tailored width and porosity distributions. Although the present formulation assumes a prescribed magnetization profile, as appropriate for hard magnetic elastomers~\citep{wu2026review}, extending it to field-dependent materials with more complex magneto-mechanical coupling represents a promising direction for future work. Overall, this work advances inverse design from simply constructing a target shape toward identifying which geometries are physically realizable, providing a physically interpretable foundation for programmable soft actuators and architected functional materials.

\section*{CRediT authorship contribution statement}
\textbf{Jiahao Li}: Writing – Original draft, Methodology, Software, Formal analysis, Investigation, Data curation, Validation.
\textbf{Yingchao Zhang}: Writing – Original draft, Methodology, Software, Formal analysis, Investigation, Data curation, Validation. 
\textbf{Weicheng Huang}: Writing – Original draft, Methodology, Software, Formal analysis, Investigation, Data curation, Validation. 
\textbf{Shenghao Ye}: Formal analysis, Investigation.
\textbf{HengAn Wu}: Writing – Review \& Editing, Methodology, Formal analysis, Investigation, Validation, Supervision, Funding acquisition, Conceptualization.
\textbf{Dominic Vella}: Writing – Review \& Editing, Methodology, Formal analysis, Investigation, Project administration, Funding acquisition, Conceptualization.
\textbf{Mingchao Liu}: Writing – Review \& Editing, Methodology, Formal analysis, Investigation, Validation, Supervision, Project administration, Funding acquisition, Conceptualization.

\section*{Data and code availability}
The datasets and code generated during the current study are available from the corresponding author upon reasonable request.

\section*{Declaration of Generative AI and AI-assisted technologies in the writing process}
During the preparation of this work, the authors used OpenAI’s ChatGPT to improve the language and readability of the text. After using this tool, the authors reviewed and edited the content as needed and take full responsibility for its publication.

\section*{Declaration of competing interest}
The authors declare that they have no known competing financial interests or personal relationships that could have appeared to influence the work reported in this paper.

\section*{Acknowledgments}
The research leading to these results has received funding from the start-up funding from Newcastle University, UK (W.H.), the Leverhulme Trust through a Philip Leverhulme Prize (D.V.), the Royal Society through a Newton International Fellowship (M.L.), and the start-up funding from the University of Birmingham (M.L.). The numerical calculations were performed on the supercomputing system in Hefei Advanced Computing Center and the Supercomputing Center of University of Science and Technology of China.

\appendix
\renewcommand{\theequation}{\Alph{section}.\arabic{equation}}

\section{Validity of the inextensibility assumption by dimensional analysis}
\label{app:C}

The model presented in this work assumes that the centerline of the magneto-elastica is inextensible, so that deformations are accommodated entirely through bending. Although this assumption has been widely adopted in previous studies irrespective of the boundary conditions~\citep{domokos1997constrained,holmes1999constrained}, it might prompt questions under highly constrained clamped--clamped conditions, where the initial geometric shape and the magnitude of the applied magnetic torques could seemingly induce stretching. Previous work for the classical elastica has shown that the arch behaves as essentially inextensible provided that it is sufficiently slender \citep{neukirch2012vibrations,pandey2014dynamics}. To justify its validity in this case, we provide a rigorous dimensional argument below, demonstrating that the inextensibility condition is a direct consequence of the rod's slender geometry, rendering it robustly independent of the boundary constraints, initial configuration, and magnetic actuation intensity.

Consider a slender rod of length $L$ with a rectangular cross-section of width $b$ and thickness $h$ ($h \ll L$), characterized by Young's modulus $E$. The structural stiffness scaling for axial stretching and bending can be expressed respectively as:
\begin{equation}
k_{\text{stretch}} \sim \frac{EA}{L} \sim \frac{Ebh}{L}
\end{equation}
\begin{equation}
k_{\text{bend}} \sim \frac{EI}{L^3} \sim \frac{Ebh^3}{L^3}
\end{equation}
where $A = bh$ is the cross-sectional area and $I = \frac{1}{12}bh^3$ is the second moment of area. Taking the  ratio of these two stiffnesses yields the dimensionless parameter:
\begin{equation}
\frac{k_{\text{bend}}}{k_{\text{stretch}}} \sim \left(\frac{h}{L}\right)^2
\end{equation}

Given the slender geometry of the rod ($h/L \ll 1$), this ratio is extraordinarily small, evaluating to $\mathcal{O}(10^{-3})$ even for a conservative slenderness ratio of $1:20$. This implies that the energy penalty for stretching the centerline is orders of magnitude higher than that for bending. From an energetic standpoint, when the clamped--clamped rod undergoes large geometric deflections, such as snapping from an initial configuration within a fixed immovable span less than $L$, the system dynamically seeks a minimum energy pathway. Rather than compressing or stretching its centerline to accommodate the boundary constraints, which would trigger the immense stretching stiffness $k_{\text{stretch}}$, the rod preferentially undergoes large-rotation, large-displacement configurations through purely bending modes, such as asymmetric snapping or higher-order buckling~\citep{wang2024transient}. 

Furthermore, this mechanism remains robust against variations in the magnitude of the magnetic actuation. Crucially, the distributed magnetic torques act equivalently as pure bending loads rather than axial forces as shown in Eq.~\eqref{eq:1}. As a result, the magnetic actuation exclusively excites the bending modes of the rod, completely bypassing the high-energy stretching modes. Consequently, while the specific deformation profiles are modulated by the initial shape and the magnitude of the magnetic torques, the material experience remains strictly within the \textit{small strain} regime. The centerline inextensibility assumption ($\epsilon_0 = 0$) therefore provides an exceptionally accurate and mathematically decoupled description of the large-deformation behavior of the magneto-elastica.

\section{Experiment settings: Fabrication of magneto-elastica}
\label{app:A}

The fabrication of the magneto-elastica follows a procedure similar to that reported in our recent works \citep{zhang2025achieving} and is briefly summarized here. A thermoplastic elastomer (SEBS) was dissolved in toluene and mixed with hard ferromagnetic particles (average diameter $5\mathrm{\mu m}$) at a weight ratio of 1:2. The Young’s modulus of the resulting composite was determined from uniaxial tensile tests to be $E=4.5 \mathrm{~MPa}$. The sample thickness was measured to be $0.4 \mathrm{~mm}$. A laser cutter was used to define the desired sample width or to introduce porous structures. Pre-magnetization along the longitudinal direction was performed under a strong magnetic field ($\sim$ 4 T), resulting in a remnant magnetization  $M=150 \mathrm{~kA/m}$, evaluated as a volume average. Deformation of the magneto-elastica was actuated using a custom-built magnetic field setup, with the field strength monitored by a Gaussmeter.

\section{Discrete Elastic Rod algorithm }
\label{app:B}

The planar discrete elastica rod (DER) approach is employed to validate the inverse design theory of magneto elastica~\citep{bergou2008discrete,jawed2018primer,huang2025tutorial}. A slender rod is represented by $N$ discrete nodes $\{ \boldsymbol{x}_0, \boldsymbol{x}_1, \ldots, \boldsymbol{x}_{N-1} \}$, which define $N-1$ edges $\boldsymbol{e}^i = \boldsymbol{x}_{i+1} - \boldsymbol{x}_i$ for $i = 0, \ldots, N-2$. Subscripts denote node-based quantities, whereas superscripts indicate edge-based quantities. Each edge $\boldsymbol{e}^i$ is associated with a material frame $\{\boldsymbol{m}^i, \boldsymbol{t}^i \}$, $\boldsymbol{m}^i$ is the normal vector and $\boldsymbol{t}^i$ is the tangent vector $\boldsymbol{t}^i = \boldsymbol{e}^i / \| \boldsymbol{e}^i \|$. The system configuration is described by the generalized coordinate vector
\[
\boldsymbol{q} =
[\boldsymbol{x}_0, \boldsymbol{x}_1, \ldots,
\boldsymbol{x}_{N-2}, \boldsymbol{x}_{N-1}]^T \in \mathcal{R}^{2N \times 1 },
\] 
which contains $2N$ degrees of freedom (DOF), where $(\cdot)^T$ denotes the transpose operator.
The stretching strain for each edge is
\begin{equation}
\varepsilon^i = \frac{\| \boldsymbol{e}^i \|}{\| \bar{\boldsymbol{e}}^i \|} - 1,
\end{equation}
where $\| \cdot \|$ denotes the Euclidean norm and $\bar{\boldsymbol{e}}^i$ is the corresponding reference edge vector.
We next formulate the bending curvature.
The discrete curvature is based on the turning angle between the two consecutive edges,
\begin{equation}
\kappa_i = 2 \frac {\tan(\phi_i / 2)} {\Delta l_{i}},
\end{equation}
where $\phi_i$ is the turning angle and $\Delta l_{i}$ is its voronoi length.
The total elastic energy is the sum of the stretching and bending,
\begin{equation}
E_s =
\frac{1}{2} \sum_{i=0}^{N-2}
EA^{i} (\varepsilon^i)^2 \| \bar{\boldsymbol{e}}^i \|.
\end{equation}
\begin{equation}
E_b =
\frac{1}{2} \sum_{i=1}^{N-2}
EI_i (\kappa_i )^2 \Delta l_i,
\end{equation}
where $EA^{i} $  (respectively $EI_i$) is the stretching (respectively bending) stiffness  of each element, which is obtained from the theoretical width distribution.
Next, we use the Helmholtz free energy for the ideal hard-magnetic soft materials. The magnetization of the $i$-th segment is assumed to be $\bm{\mathcal{M}}^i$, and the external magnetic field applied to this segment is $\mathbf{B}_a^i$.
We assume the magnetized moments will not affect the magnetic field during actuation.
The discrete form of the Helmholtz free energy can be written as: 
\begin{equation}
E_{m}=\sum_{i=0}^{N-2}E_{m}^i=-\sum_{i=0}^{N-2}(\bm{\mathcal{M}}^i\cdot\mathbf{B}_a^i) \; ||\bar{\mathbf{e}}^i||,
\end{equation}
where the magnetization vector of $i$-th edge is:
\begin{equation}
    \bm{\mathcal{M}}^i=\frac{A^{i}}{\mu_0}\left[(\mathbf{m}^i\otimes\bar{\mathbf{m}}^i)+(\mathbf{t}^i\otimes\bar{\mathbf{t}}^i)\right]\cdot\mathbf{B}_r^i,
\end{equation}
where $\otimes$ is the tensor product, $A^{i}$ is the local cross-section area,  
 $\left[(\mathbf{m}^i\otimes\bar{\mathbf{m}}^i)+(\mathbf{t}^i\otimes\bar{\mathbf{t}}^i)\right]$ is the reduced deformation gradient of a 1D rod, $\mu_0$ is the magnetic permeability of free space, and $\mathbf{B}^{i}_{r}$ is the local remnant magnetic flux density.
Finally, we use the dynamic relaxation method to solve the nonlinear magnetic-elastic rod system.
The update from time $t_k$ to $t_{k+1} = t_k + h$ satisfies
\begin{equation}
\mathbb{M}
(\boldsymbol{q}_{k+1} - \boldsymbol{q}_k - h \boldsymbol{v}_k)
-
h^2 (\boldsymbol{F}_{k+1}^{\mathrm{int}}
+ \boldsymbol{F}_{k+1}^{\mathrm{mag}}
+ \boldsymbol{F}_{k+1}^{\mathrm{ext}})
= \boldsymbol{0},
\label{eqn:newton}
\end{equation}
where $\boldsymbol{v} = \dot{\boldsymbol{q}}$ is the velocity of the DOF vector, $\boldsymbol{F}^{\mathrm{int}}$ is the internal elastic force, $\boldsymbol{F}^{\mathrm{mag}}$ is the magnetic actuated force, 
 $\boldsymbol{F}^{\mathrm{ext}}$ is the other external forces, and $\mathbb{M}$ is the lumped mass matrix.
The Jacobian matrix associated with Eq.~\eqref{eqn:newton} is
\begin{equation}
\mathbb{J}
=
\mathbb{M}
+
h^2
\left( \frac{\partial^2 (E_s + E_b + E_{m})}
{\partial \mathbf{q}^{2}} \right),
\end{equation}
The Jacobian matrix is banded, and the resulting linear system can be solved with computational complexity $\mathcal{O}(N)$.


\end{document}